\documentclass{ws-ijbc}
\usepackage{ws-rotating} 
\usepackage{graphicx}
\usepackage{epstopdf}

\usepackage[usenames,dvipsnames]{pstricks}
\usepackage{epsfig}
\usepackage{pst-grad} 
\usepackage{pst-plot} 

\begin{document}

\catchline{}{}{}{}{} 

\markboth{Chris Antonopoulos et. al}{Linear And Nonlinear Arabesques: A Study Of Closed Chains Of Negative 2-Element Circuits}

\title{LINEAR AND NONLINEAR ARABESQUES: A STUDY OF CLOSED CHAINS OF NEGATIVE 2-ELEMENT CIRCUITS}

\author{CHRIS ANTONOPOULOS}
\address{Department of Physics, University of Aberdeen, Institute for Complex Systems and Mathematical Biology, SUPA,\\Aberdeen, AB24 3UE, United Kingdom\\chris.antonopoulos@abdn.ac.uk}

\author{VASILEIOS BASIOS}
\address{Service de Physique des Syst\`{e}mes Complexes et M\'{e}canique Statistique, Universit\'{e} Libre de Bruxelles, Interdisciplinary Center for Nonlinear Phenomena and Complex Systems (CeNoLi),\\Brussels, CP231, Belgium\\vbasios@ulb.ac.be}

\author{JACQUES DEMONGEOT}
\address{Faculty of Medicine of Grenoble, University Joseph Fourier, Laboratory AGIM (Age, Imaging \& Modelling) CNRS,\\La Tronche, FRE 3405, France\\jacques.demongeot@yahoo.fr}

\author{PASQUALE NARDONE}
\address{D\'{e}partement de Physique, Universit\'{e} Libre de Bruxelles, Service de Physique G\'{e}n\'{e}rale, Laboratoire de Didactique des Sciences Physiques,\\Brussels, CP238, Belgium\\pnardon@ulb.ac.be}

\author{REN\'{E} THOMAS}
\address{Service de Chimie Physique et Biologie Th\'{e}orique, Universit\'{e} Libre de Bruxelles,\\Brussels, CP231, Belgium\\ingathomas@skynet.be}

\maketitle

\begin{history}
\received{(to be inserted by publisher)}
\end{history}

\begin{abstract}
In this paper we consider a family of dynamical systems that we call ``arabesques'', defined as closed chains of 2-element negative circuits. An $n$-dimensional arabesque system has $n$ 2-element circuits, but in addition, it displays by construction, two $n$-element circuits which are both positive vs one positive and one negative, depending on the parity (even or odd) of the dimension $n$. In view of the absence of diagonal terms in their Jacobian matrices, all these dynamical systems are conservative and consequently, they can not possess any attractor. First, we analyze a linear variant of them which we call ``arabesque 0'' or for short ``A0''. For increasing dimensions, the trajectories are increasingly complex open tori. Next, we inserted a single cubic nonlinearity that does not affect the signs of its circuits (that we call ``arabesque 1'' or for short ``A1''). These systems have three steady states, whatever the dimension is, in agreement with the order of the nonlinearity. All three are unstable, 
as there can not be any attractor in their state-space. The 3D variant (that we call for short ``A1\_3D'') has been analyzed in some detail and found to display a complex mixed set of quasi-periodic and chaotic trajectories. Inserting $n$ cubic nonlinearities (one per equation) in the same way as above, we generate systems ``A2\_$n$D''. A2\_3D behaves essentially as A1\_3D, in agreement with the fact that the signs of the circuits remain identical. A2\_4D, as well as other arabesque systems with even dimension, has two positive $n$-circuits and nine steady states. Finally, we investigate and compare the complex dynamics of this family of systems in terms of their symmetries.
\end{abstract}


\section{Introduction}

The fact that negative retroaction is involved in periodicity and homeostasis was already clear in the works by C. Bernard \cite{Bernard1943} who called it ``elasticity'' and W. B. Cannon \cite{Cannon1932} who invented the term ``homeostasis''. The notion that positive retroaction is involved in multistationarity as the occurrence of more than one steady state, either stable or not, is much more recent, however. A basis for the present understanding of these ``nontrivial'' processes is the recognition that, among the terms of the Jacobian matrix (or more generally in the graph of interactions), only those terms that belong to a circuit are present, by construction, in the characteristic equation of the system, and hence, only these terms are involved in the kind of the steady states. By the term circuits we refer to those sets of terms of the Jacobian matrix of the dynamical system whose row and column indices are in circular permutation (see \cite{EisenfeldanddeLisi} who use a different terminology and 
\cite{Thomas1999,ThomasandKaufman2001}). Circuits are positive or negative according to the sign of the product of their terms.

The functions of positive and negative circuits are not only distinct, but contrasting. It has become clear that the presence of a positive circuit in the Jacobian matrix or in the graph of interactions of a system, is a necessary condition for multistationarity (see conjecture by \cite{Thomas1981}, formal demonstrations in \cite{Plahte1995,Snoussi1998} for systems with constant sign partial derivatives and \cite{Soule2003} for the general case). It is also widely recognized that the presence of a 2- or more-elements negative circuit is a necessary condition for a persistent periodicity (conjecture by \cite{Thomas1981} which is demonstrated, so far limited to discrete systems, in \cite{Richard2011}). Concerning the number and nature of attractors in relation to circuits in systems described by differential equations, one can see for example \cite{Tsypkin1958,Bohn1961,Richardetal2011}, and for Boolean networks \cite{Goles1985,Aracenaetal2004A,Aracenaetal2004B,Demongeotetal2012A}).

In this paper, we investigate the behavior of conservative dynamical systems that we call ``arabesques'' whose simple form can be described, through the structure of their Jacobian matrix $J$, as closed chains of three or more (depending on the dimension $n$) 2-element negative circuits. An $n$ dimensional arabesque system has $n$ negative $2$-element circuits, but in addition it displays by construction, two $n$-element circuits both positive vs one positive and one negative, depending on the parity (even or odd) of the dimension $n$. However, they can never be both negative.

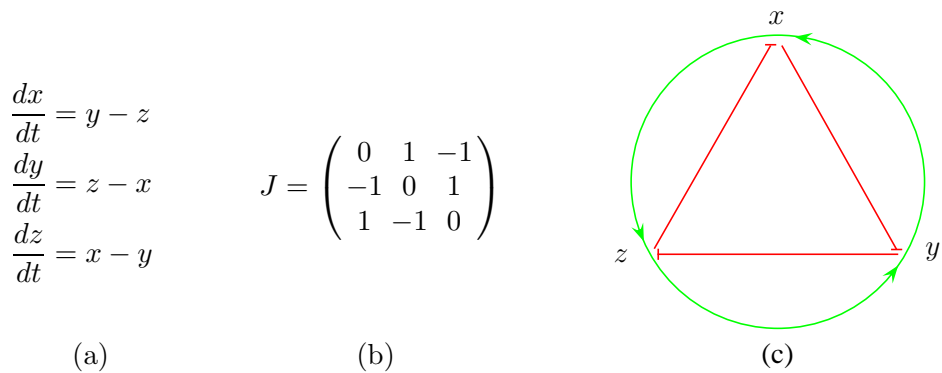
\begin{figure}[!ht]
\centering
\begin{minipage}{0.3\textwidth}
 \begin{eqnarray*}
 \frac{dx}{dt} & = & y-z\\
 \frac{dy}{dt} & = & z-x\\
 \frac{dz}{dt} & = & x-y
 \end{eqnarray*}
\rput(2.8,-0.58){(a)}
 \label{fig:figure}
\end{minipage}
\hspace{-1.5cm}
\begin{minipage}{0.3\textwidth}
\begin{eqnarray*}
J=\left(\begin{array}{ccc}
0 & 1 & -1\\
-1 & 0 & 1\\
1 & -1 & 0
\end{array}\right)
\end{eqnarray*}
\rput(2.5,-1.3){(b)}
\end{minipage}
\begin{minipage}{0.3\textwidth}
\scalebox{0.53} 
{
\begin{pspicture}(0,-4.606094)(8.817813,4.606094)
\pscircle[linewidth=0.04,linecolor=green,dimen=outer](4.363125,0.16359375){3.7}
\psline[linewidth=0.04cm,linecolor=red](1.263125,-1.5164063)(4.183125,3.5835938)
\psline[linewidth=0.04cm,linecolor=red](4.463125,3.5835938)(7.343125,-1.5364063)
\psline[linewidth=0.04cm,linecolor=red](1.343125,-1.6564063)(7.383125,-1.6564063)
\psline[linewidth=0.04cm,linecolor=red](7.203125,-1.5364063)(7.483125,-1.5364063)
\psline[linewidth=0.04cm,linecolor=red](4.043125,3.6035938)(4.323125,3.6035938)
\psline[linewidth=0.04cm,linecolor=red](1.363125,-1.5164063)(1.363125,-1.7964063)
\usefont{T1}{ptm}{m}{n}
\rput(4.3320312,4.273594){\huge $x$}
\usefont{T1}{ptm}{m}{n}
\rput(0.43203124,-1.6864063){\huge $z$}
\usefont{T1}{ptm}{m}{n}
\rput(8.252031,-1.5864062){\huge $y$}
\psline[linewidth=0.134cm,linecolor=green,arrowsize=0.05291667cm 2.0,arrowlength=1.4,arrowinset=0.4]{<-}(1.003125,-1.3364062)(0.883125,-1.0364063)
\psline[linewidth=0.134cm,linecolor=green,arrowsize=0.05291667cm 2.0,arrowlength=1.4,arrowinset=0.4]{<-}(7.383125,-1.9364063)(7.203125,-2.1764061)
\psline[linewidth=0.134cm,linecolor=green,arrowsize=0.05291667cm 2.0,arrowlength=1.4,arrowinset=0.4]{<-}(4.783125,3.8035936)(5.083125,3.7635937)
\usefont{T1}{ptm}{m}{n}
\rput(4.3632812,-4.206406){\huge (c)}
\end{pspicture} 
}
\end{minipage}
\caption{The equations of motion in panel (a), the Jacobian matrix $J$ in panel (b) and the graph in panel (c) of the linear variant A0 for $n=3$. In panel (c), the positive and negative interactions are drawn in green and red respectively.\label{fig:1-1}}
\end{figure}

It is important to emphasize that there are no diagonal terms in their Jacobian matrices (equivalently, there are no 1-element circuits) and hence these systems are conservative and thus their state-space volume is conserved ($\nabla{}_{x}\frac{d{\vec{x}}}{dt}=0$), where $\vec{x}$ is a shorthand for the $n$ dimensional vector of states $\{x_{k}\}_{k=1}^{n}$. In view of the absence of any diagonal element in their Jacobian matrix, one does not expect the presence of any attractor.

These systems, of which the linear variant is briefly described in \cite{Thomasetal2006}, are built on the basis of 2-element negative circuits between variables $x_{k+1}$ and $x_{k-1}$ as follows: $\frac{dx_{k}}{dt}=x_{k+1}-x_{k-1}$ where $k=1,2,3,\ldots,n$ follow a circular permutation. For convenience, we will call variables $x,y,z,\ldots$ as $x_{1},x_{2},x_{3},\ldots$ etc..

Matrix $J$ in panel (b) of Fig. \ref{fig:1-1} permits to read directly the circuits of the $3$-variable linear system A0: it has three 2-element negative circuits $\{\{J_{12},\;J_{21}\},\;\{J_{13},\;J_{31}\},\;\{J_{23},J_{32}\}\}$ and two 3-element circuits, one positive $\{\{J_{12},\;J_{23},\;J_{31}\}\}$ and one negative $\{\{J_{13},\;J_{21},\;J_{32}\}\}$ where $J_{ij}$ denotes the $(i,j)$th entry of $J$. More generally, whatever $n$, this construction implies, in addition to the presence of $n$ 2-element negative circuits, the presence of two $n$-element circuits, either one positive and one negative, or both positive, depending on whether $n$ is odd or even.

We would like to mention that we use, throughout the paper, the following convention for the signs of the eigenvalues of $J$. The signs of real eigenvalues can be positive, negative or zero according to the case at hand. Regarding the pairs of conjugated complex eigenvalues, we list behind a ``$/$'' the signs of their real part. For example, in a 3D system, the symbol $\{++-\}$ means that all three eigenvalues are real and that two of them are positive (a saddle-node), while ${\{+/--}\}$ means that there is a real positive eigenvalue and a pair of conjugated eigenvalues whose real part is negative (one of the two types of saddle-foci) and finally, $\{+/0\;0\}$ means that there is a positive real eigenvalue and that the pair of conjugated complex eigenvalues are purely imaginary, i.e. their real parts are zero.

We now pass to the next section where we present and discuss about methodologies that enable one to characterize efficiently and fast the dynamical nature of a given trajectory (i.e. discriminate between order and chaos) of a dynamical system, such as those we study herein.

\section{Methods for discriminating regular from chaotic motion}

In the theory of dynamical systems, the discrimination between regular and chaotic motion is of crucial importance since in ordered domains we have predictability while in chaotic regions, after a time period, we are unable to predict the evolution of the system due to its sensitive dependence on neighboring initial conditions, a \textit{conditio sine qua non} signature of chaos \cite{Lichtenbergetal1983,Bountisetal2012}. This is a characteristic and important feature found in many applied sciences such as Physics, Chemistry, Biology and Engineering etc., where the dynamics is often dissipative and the motion occurs on a strange attractor \cite{Sparrow1982,Lichtenbergetal1983}.

However, the ability to distinguish between order and chaos becomes much more difficult when the motion is conservative and the number of equations of motion of the system (or its dimension) is increasing, mainly since we can not visualize their state or phase-space. It naturally follows that we need methods that can help us decide accurately, fast and efficiently about the ordered or chaotic character of a given trajectory \cite{Bountisetal2012}. Many methods have been developed over the years aiming to tackle this problem. The oldest of all is the well known method of the Poincar\'{e} surface of section (PSS) (see \cite{Lichtenbergetal1983}) which takes advantage of the intersections of a trajectory with an $m$D ``surface'' ($m$ being its dimension) that is transversal to the flow generated by the system. It can be used mainly in conservative dynamical systems with a small number of equations of motion, such as for example the 3D version of the arabesque systems we consider in this paper. For example, in 2 
degrees of freedom (DOF) Hamiltonian systems where the phase-space is 4D, the energy integral can be used to lower the dimension to 3, thus rendering the PSS into a 2D surface (i.e. into a plane). However, the results of this method are difficult to interpret in systems with more than 2 DOF or in many dimensions in general.

There have also been developed over the past years a great variety of methods aiming to face this subtle problem \cite{Skokos2010}, however we prefer to stick for the present study to an efficient methodology due to its success to discriminate between order and chaos. In particular, in \cite{Skokos2001,Skokosetal2003,Skokosetal2004,Antonopoulosetal2006A,Manosetal2008,Manosetal2011,Bountisetal2012}, the authors have used a fast, accurate and easy to implement and compute method to distinguish regular from chaotic motion in various conservative dynamical systems called the ``Smaller Alignment Index'' (SALI) method. In \cite{Skokosetal2007}, it is introduced and used its generalization, called the ``Generalized Alignment Index'' (GALI) and in \cite{Antonopoulosetal2006C} a faster implementation is reported, called the ``Linear Dependence Index'' (LDI), particularly suited for multidimensional conservative systems. We will briefly recall the definition of SALI in Subsec. \ref{sub:The-SALI-method} and show its 
effectiveness in distinguishing regular from chaotic motion in the class of low dimensional arabesque systems. Before that however, in the next Subsec. \ref{sub:The-Lyapunov-exponents}, we prefer to make a short introduction to the theory of Lyapunov exponents that we will utilize later in the paper, since it is a standard and well accepted methodology for tackling the same kinds of problems.

\subsection{The Lyapunov exponents\label{sub:The-Lyapunov-exponents}}

As we have already pointed out, knowing whether the trajectories of a dynamical system are ordered or chaotic is fundamental for understanding its behavior. In the dissipative case, this distinction is easily made as both types of motion are finally attracting. In conservative systems, however, the distinction between order and chaos is often a delicate issue, for example when the ordered and/or chaotic regions are small-sized and interwined, especially in multi-dimensional systems where it is not tractable to visualize their dynamics. For all these, it becomes crucial to make use of quantities that can decide if a trajectory is ordered or chaotic, independently of the dimension of the state or phase-space and/or our ability to inspect visually the dynamics.

The well-known and commonly used traditional method for this purpose is the evaluation of the maximal Lyapunov Characteristic Exponent (LCE) $\sigma_{1}$. If $\sigma_{1}>0$ the trajectory is characterized as chaotic. The theory of Lyapunov exponents was applied to characterize chaotic trajectories by Oseledec \cite{Oseledec1968}, while the connection between Lyapunov exponents and exponential divergence of nearby trajectories was given in \cite{Benettinetal1976,Pesin1977}. Benettin et al. \cite{Benettinetal1980A} studied theoretically the problem of the computation of all LCEs (Lyapunov spectrum) and proposed in \cite{Benettinetal1980B} an algorithm for their numerical computation. In particular, $\sigma_{1}$ is computed as the limit for $t\rightarrow\infty$ of the quantity:
\begin{equation}\label{eq:sigma1}
L_{1}(t)=\frac{1}{t}\ln\frac{\|\vec{w}(t)\|}{\|\vec{w}(0)\|},\;\sigma_{1}=\lim_{t\rightarrow\infty}L_{1}(t)
\end{equation}
where $\|\cdot\|$ denotes the usual Euclidean norm of $\mathbb{R}^{n}$ and $\vec{w}(0)$, $\vec{w}(t)$ are deviation vectors from the trajectory one wants to characterize, at times $t=0$ and $t>0$ respectively. It has been shown that the above limit is finite, independent of the choice of the metric for the state-space and converges to $\sigma_{1}$ for almost all initial vectors $\vec{w}(0)$ \cite{Oseledec1968,Benettinetal1980A,Benettinetal1980B}. Similarly, all other LCEs $\sigma_{2},\;\sigma_{3}$ etc. are computed as the limits for $t\rightarrow\infty$ of some appropriate quantities, $L_{2}(t),\;L_{3}(t)$ etc. (see for example \cite{Benettinetal1980B,Skokos2010} for a detailed analysis). We note that, throughout the paper, whenever we need to compute the values of the LCEs, we apply the numerical algorithm proposed by Benettin et al. \cite{Benettinetal1980B}.

The time evolution of the deviation vectors is given by solving the variational equations (see Subsec. \ref{sub:The-SALI-method} and \cite{Skokos2010} for more details). In general, for almost any choice of a sufficiently small initial deviation vector $\vec{w}(0)$, the limit for $t\rightarrow\infty$ of Eq. \eqref{eq:sigma1} gives the same $\sigma_{1}$. For reasons of avoiding numerical overflows, since the exponential growth of $\vec{w}(t)$ occurs for short time intervals, one has to stop its evolution after some time $T_{1}$, and record the computed $L_{1}(T_{1})$, normalize vector $\vec{w}(t)$ and repeat the calculation for the next time interval $T_{2}$, etc. obtaining finally $\sigma_{1}$ as an average over many time intervals $T_{i},\;i=1,\ldots,N$ as:
\begin{equation}
\sigma_{1}=\frac{1}{N}\sum_{i=1}^{N}L_{1}(T_{i})\nonumber.
\end{equation}
Since $\sigma_{1}$ is influenced by the evolution of $\vec{w}(0)$, the time needed for $L_{1}(T)$ (or $L_{1}(T_{i})$) to converge is not known a priori and may become extremely long. This makes it often difficult to infer whether $\sigma_{1}$ finally tends to a positive value (which signifies a chaotic behavior) or converges to zero (which signifies a regular motion).

In the recent years, several methods have been introduced, which try to avoid this problem by studying the evolution of deviation vectors. In the present paper, as we have already mentioned, we will make use of the SALI method, first introduced in \cite{Skokos2001}, that we will describe in the following Subsec. \ref{sub:The-SALI-method}. It has been applied successfully to several multi-dimensional maps \cite{Skokosetal2002} and Hamiltonian systems \cite{Skokosetal2003,Skokosetal2004,Antonopoulosetal2006A,Manosetal2008,Skokos2010,Manosetal2011} and was found to converge rapidly to zero for chaotic trajectories, while exhibits small fluctuations around non-zero values for ordered trajectories. It is exactly this different behavior of SALI which makes it an ideal indicator of chaoticity: Unlike the maximal LCE $\sigma_{1}$, as we will see in the next subsection, SALI does not start at every time step a new calculation of the deviation vectors, but takes into account information about their convergence on the 
unstable manifold 
from all previous times.

\subsection{The Smaller Alignment Index\label{sub:The-SALI-method} }

Let us consider the $n$D state-space of an autonomous conservative dynamical system:
\begin{equation}
\frac{d\vec{x}(t)}{dt}=\vec{\Phi}(\vec{x}(t))\label{eq:dyn_sys_SALI}
\end{equation}
where $\vec{x}(t)=(x_{1}(t),x_{2}(t),\ldots,x_{n}(t))$ and $\vec{\Phi}(\vec{x}(t))=(\Phi_{1}(\vec{x}(t)),\Phi_{2}(\vec{x}(t)),\dots,\Phi_{n}(\vec{x}(t)))$. The time evolution of a trajectory associated with the initial state $\vec{x}(t_{0})$ at initial time $t_{0}$ is defined as the solution of the system of $n$ first order differential equations (ODEs) \eqref{eq:dyn_sys_SALI}.

In order to define SALI we need to introduce the notion of variational equations (see for example \cite{Skokos2010}). These equations are the corresponding linearized equations of the system of ODEs \eqref{eq:dyn_sys_SALI} (i.e. given by the Jacobian matrix $J$ of $\vec{\Phi}(\vec{x}(t))$), about a reference trajectory $\vec{x(t)}$ and are defined by the relation:
\begin{equation}
\frac{d\vec{w}_{i}(t)}{dt}=J(\vec{x}(t))\vec{w_{i}}(t),\;i=1,2\label{eq:jac_sys_SALI}
\end{equation}
where $J(\vec{x}(t))$ is the Jacobian matrix of the right hand side of the system of ODEs \eqref{eq:dyn_sys_SALI} calculated about the trajectory $\vec{x}(t)=(x_{1}(t),x_{2}(t),\ldots,x_{n}(t))$. $\vec{w}_{1}(t)$ and $\vec{w}_{2}(t)$ are known as deviation vectors because they describe small deviations of neighbor trajectories from $\vec{x}(t)$ (see Eq. \eqref{eq:deviationvectors_components}). We define the two Alignment Indices (ALI) as:
\begin{equation}
\mathrm{ALI_{-}}(t)=\Biggl\|\frac{\vec{w}_{1}(t)}{\|\vec{w}_{1}(t)\|}-\frac{\vec{w}_{2}(t)}{\|\vec{w}_{2}(t)\|}\Biggr\|\label{eq:ALIminus}
\end{equation}
and
\begin{equation}
\mathrm{ALI_{+}}(t)=\Biggl\|\frac{\vec{w}_{1}(t)}{\|\vec{w}_{1}(t)\|}+\frac{\vec{w}_{2}(t)}{\|\vec{w}_{2}(t)\|}\Biggr\|.\label{eq:ALIplus}
\end{equation}
SALI is then defined as:
\begin{equation}
\mathrm{SALI}(t)=\min(\mathrm{ALI_{-}}(t),\mathrm{\;ALI_{+}}(t)).\label{eq:def_SALI}
\end{equation}
It follows that one can normalize $\vec{w}_{1}(t)$ and $\vec{w}_{2}(t)$ at each integration time step without affecting their inner angle. It comes out that when $\mathrm{ALI_{-}(t)\rightarrow0}$, Eqs. \eqref{eq:ALIminus} and \eqref{eq:ALIplus} imply that the two deviation vectors tend to become collinear and parallel because their inner angle tends to 0 and so $\mathrm{ALI_{+}(t)\rightarrow2}$, while when $\mathrm{ALI_{+}(t)\rightarrow0}$ the two deviation vectors tend to become collinear and anti-parallel because their angle tends to $\pi$ and so $\mathrm{ALI_{-}(t)\rightarrow2}$.

Thus, if the trajectory under consideration is chaotic then from Eq. \eqref{eq:def_SALI} we get: $$\lim_{t\rightarrow\infty}\mathrm{SALI(t)=\min(0,2)=0}$$ because both deviation vector directions tend to coincide with the direction of the most unstable neighbor manifold of the variational equations as $t\rightarrow\infty$ \cite{Voglisetal1999,Skokos2001}. This implies that the two deviation vectors tend to become collinear and either coincide or become anti-parallel to each other.

On the other hand, if the trajectory is regular (e.g. quasi-periodic) then SALI fluctuates around non-zero positive 
values because both deviation vectors tend to become tangential to the torus on which the trajectory is confined. 
The reason is that $\vec{w}_{1}(t)$ and $\vec{w}_{2}(t)$ have components evolving along and across the torus 
in directions given by two independent vector fields: the Hamiltonian and a second local quasi-integral of the 
motion \cite{Skokosetal2003}.

This distinct behavior of SALI for regular and chaotic trajectories makes it a very clear and reliable way to distinguish between order and chaos in conservative dynamical systems. It is evident that the choice of $\vec{w}_{1}(t_{0})$ and $\vec{w}_{2}(t_{0})$ is arbitrary and does not affect the results of the method. This is additionally ensured by results presented in \cite{Skokos2001,Skokosetal2003}. Thus, along this direction, we adopt and use throughout this paper as deviation vectors:
\begin{equation}
\begin{array}{c}
\vec{w}_{1}(0)=(1,0,0,\ldots,0),\\
\vec{w}_{2}(0)=(0,1,0,\ldots,0).
\end{array}\label{eq:deviationvectors_components}
\end{equation}

\subsection{Circuits and Frontiers: Partition of state-space into domains according to the nature of the eigenvalues of the Jacobian matrix}\label{sub:=00201CFrontiers=00201D:-Partition-of}

The state-space of a dynamical system can be partitioned into domains according to the eigenvalues of its Jacobian matrix $J$: their sign if they are real, complex or purely imaginary, etc. Since the eigenvalues in a domain of the state-space determine the precise nature of any steady state present in it, this approach can provide a global view of the state-space of the system \cite{ThomasandKaufman2001,Thomasetal2005}\footnote{This section represents a brief actualization of the more detailed papers mentioned. We take this opportunity to correct a printing mistake found in Appendix 3 of \cite{ThomasandNardone2009}: in the symbolic calculation program ``To nDimensions (For Frontiers $F_{1}$, $F_{2}$ and $F_{4}$ only)'' one should read ``eqcar = Collect{[}Det{[}- Outer{[}D, syst,var{]} + λ IdentityMatrix{[}dim{]}{]}/.values, λ{]}; (a minus sign ``-'' was missing).}.

The characteristic equation, $\det[J-\lambda I_{n}]=0$ (where $\det[\cdot]$ is the determinant of the argument and $I_{n}$ is the identity matrix in $n$ dimensions), whose roots are the $n$ eigenvalues of $J$, can be written equivalently, using the two standard forms as:
\begin{equation}
\lambda^{n}+a_{n-1}\lambda^{n-1}+a_{n-2}\lambda^{n-2}+\ldots+a_{1}\lambda+a_{0}=0\label{eq:CharEq1}
\end{equation}
and
\begin{equation}
(\lambda-\lambda_{1})(\lambda-\lambda_{2})\ldots(\lambda-\lambda_{n-1})(\lambda-\lambda_{n})=0.\label{eq:CharEq2}
\end{equation}
Coefficients $a_{n-i},\;i=1,\ldots,n$ of the characteristic polynomial (which results from the expansion of the determinant), are, respectively, the sums of all the cyclic permutations (with the appropriate signatures, see below), of $1,2,\ldots,n$ elements of $J$. In other words, they are the sums of the values of the circuits and unions of disjoint circuits involving $1,2,\ldots,n$ variables, respectively, each with a sign ($+$ or $-$) that depends on the parity (even or odd) of the number of circuits involved in the union of disjoint circuits considered. 

The value of a circuit is the product of its elements. For the sake of clarity, we list the elements of a circuit with a simple space (implicit multiplication) but use a (multiplicative) dot to separate unions of disjoint circuits. For example, $(a_{11}\cdot a_{23}a_{32})$ represent the union of the 1-element circuit $a_{11}$ and the 2-element circuit $a_{23}a_{32}$.

Consequently, coefficient $a_{n-1}$ represents the sum of the values of the 1-element circuits, i.e. the diagonal terms of $J$, in three dimensions: $-(a_{11}+a_{22}+a_{33})$. Coefficient $a_{n-2}$ represents the sum of the values of the 2-circuits, i.e. in 3D it is $-(a_{12}a_{21}+a_{23}a_{32}+a_{31}a_{13})$ and of the unions of two disjoint 1-circuits, $+(a_{11}\cdot a_{22}+a_{22}\cdot a_{33}+a_{33}\cdot a_{11})$. Coefficient $a_{n-3}$ represents the sums of the values of the 3-circuits, $-(a_{12}a_{23}a_{31}+a_{13}a_{21}a_{32})$, of the unions of disjoints 1- and 2-circuits, $+(a_{11}\cdot a_{23}a_{32}+a_{22}\cdot aa_{31}+a_{33}\cdot a_{12}a_{21})$ and of the unions of three disjoint 1-circuits, $-(a_{11}\cdot a_{22}\cdot a_{33})$, etc..

The fact that the characteristic equation is entirely built from the circuits present in $J$ explains the crucial role exerted by circuits in the dynamics of systems. In fact, the arabesque systems treated in this work have been conceived and built in terms of circuits. It may be of interest to remark that the above permits to write the characteristic equation ``by hand'' when the number of circuits is rather limited and the multiplicity not too high.

Note that in view of Eqs. \eqref{eq:CharEq1} and \eqref{eq:CharEq2}, the first coefficient (the sum of the diagonal elements) is, apart from the sign, the sum of the eigenvalues and that the last coefficient (the determinant of $J$) equals $(-1)^{n}$ the product $\Pi$ of the eigenvalues. Product $\Pi$, coefficient $a_{0}$ and $\det[J]$ which is the determinant of the Jacobian matrix, are all related by: $a_{0}=(-1)^{n}\Pi_{i=1}^n\lambda_{i}=(-1)^{n}\det[J]$.

The partition of phase space according to the kind of eigenvalues is realized by utilizing the concept of ``frontier'' \cite{Thomasetal2005,ThomasandNardone2009}. Frontiers are defined as functions of the invariants of the characteristic equation of the local Jacobian matrix. The detailed algebraic derivation by a symbolic calculation program can be found in Appendix 3 of \cite{ThomasandNardone2009}. It readily translates these general operations into defined analytic expressions according to the nature of each system. Among the frontiers described in \cite{ThomasandNardone2009}, we shall use herein only frontiers $F_{1}$, $F_{2}$, which account for changes of the sign of the eigenvalues of $J$, and $F_{4}$, which deals with the transition from real to complex eigenvalues of $J$.
\begin{itemize}
\item{$F_{1}$ is defined as the set of points where:
\begin{equation}
a_{0}=(-1)^{n}\prod_{i=1}^{n}\lambda_{i}=(-1)^{n}\det[J]=0.
\end{equation}
As $a_{0}$ equals to the product of the eigenvalues, frontier $F_{1}$ is the locus of the points of the state-space at which at least one of the eigenvalues is nil. In simple cases, one of the eigenvalues changes sign as this frontier is crossed, which then separates domains that differ by the sign of one eigenvalue.}

\item{$F_{2}$ is the locus of the points at which there exists a pair of purely imaginary eigenvalues of $J$. In simple cases, upon crossing this frontier, a pair of complex conjugated eigenvalues switches from two complex values with positive real parts (symbolized as $\{/++\}$) to two complex eigenvalues with negative real parts (symbolized as $\{/--\}$). Upon frontier $F_{2}$, one has then $\{/0\;0\}$.}

\item{Finally, we define frontier $F_{4}$ as follows: For 2D and 3D systems, this boundary was defined analytically as the discriminant of the characteristic equation of $J$, in order to discriminate real from complex eigenvalues. For higher dimensions however, the very notion of discriminant vanishes and for this reason we define $F_{4}$ to be the locus of the points where there are two equal eigenvalues. Whenever a pair of complex eigenvalues switches to a pair of real eigenvalues, we are on such a $F_{4}$ frontier.}
\end{itemize}

\subsection{Conserved quantities and symmetries}

Another aspect that we need to bring forth in focus is the conservative nature of all different types of arabesque systems considered in this paper. Thus, let us base ourselves initially on the linear arabesque system A0 and in particular in its 3D variant (see Fig. \ref{fig:1-1}) given by:
\begin{eqnarray*}
\frac{dx}{dt} & = & y-z\\
\frac{dy}{dt} & = & z-x.\\
\frac{dz}{dt} & = & x-y
\end{eqnarray*}
It is straightforward to verify and actually the same holds for the $n$D variant of the linear class, that the above system conserves a quadratic form $C$, and two functions $H_{1}$ and $H_{2}$ which are integrals of motion:
\begin{equation}
H_{1}(x,y,z)=x+y+z,\label{eq:H1}
\end{equation}
\begin{eqnarray}
H_{2}(x,y,z) & = & \frac{1}{2}(x^{2}+y^{2}+z^{2}),\label{eq:H2}
\end{eqnarray}
\begin{equation}
C(x,y,z)=\frac{1}{2}\left((y-z)^{2}+(z-x)^{2}+(x-y)^{2}\right).\label{eq:C}
\end{equation}
By the construction of the class of linear arabesque systems, functions $C,\;H_{1}$ and $H_{2}$ reduce to zero due to cyclic permutation symmetries. Simply, considering their definition and substituting by the right hand side of the equations of panel (a) of Fig. \ref{fig:1-1}, we get:
\begin{equation}
dC=\frac{\partial C}{\partial x}dx+\frac{\partial C}{\partial y}dy+\frac{\partial C}{\partial z}dz\;\Rightarrow\frac{dC}{dt}=0
\end{equation}
and
\begin{equation}
dH_{1,2}=\frac{\partial H_{1,2}}{\partial x}dx+\frac{\partial H_{1,2}}{\partial y}dy+\frac{\partial H_{1,2}}{\partial z}dz\;\Rightarrow\frac{dH_{1,2}}{dt}=0.
\end{equation}
Additionally, the same straightforward calculations hold for the new function: 
\begin{equation}
H_{a+1}(x,y,z)=\frac{1}{a+1}(x^{a+1}+y^{a+1}+z^{a+1})\label{eq:Ha_ham}
\end{equation}
where $a\neq-1$, for the system, which is also a member of the arabesque larger class of systems with all variables participating via simple power laws:
\begin{equation}
\begin{array}{c}
\frac{dx}{dt}=y^{a}-z^{a}\\
\frac{dy}{dt}=z^{a}-x^{a}.\\
\frac{dz}{dt}=x^{a}-y^{a}
\end{array}\label{eq:A1_3Da}
\end{equation}
Still, similar statements hold for this new system, as it shares the same permutation properties as the linear one (see panel (a) of Fig. \ref{fig:1-1}). Evidently, system \eqref{eq:A1_3Da} conserves the functions $C$, $H_{1}$ and $H_{a+1}$. The cyclic permutation property is responsible for the state-space conservation condition given by:
\begin{equation}
dH_{a+1}=x^{a}(y^{a}-z^{a})+y^{a}(z^{a}-x^{a})+z^{a}(x^{a}-y^{a})=0.
\end{equation}
The proof is the same as for the functions $H_{1}$ and $C$.

We would like to emphasize on two remarks here. First, one can easily show that the more general system defined by: 
\begin{eqnarray}
\frac{dx}{dt} & = & y^{k}-z-z^{k}+y\nonumber \\
\frac{dy}{dt} & = & z^{k}-x-x^{k}-z\label{eq:A1_3Dk}\\
\frac{dz}{dt} & = & x^{k}-y-y^{k}+x\nonumber 
\end{eqnarray}
still conserves function $H_{1}$, given by Eq. \eqref{eq:Ha_ham}, hence we see that an ``energy'' dissipation or conservation condition, is not merely due to the presence of a nonlinearity, but depends solely on the symmetry breaking between the nonlinearity of the variables.

\section{Study of the arabesque systems}
\subsection{Linear variant}

In this basic version of arabesque systems which we call ``arabesque 0'' or for short ``A0'' (for $n=3$ and its Jacobian matrix see Fig. \ref{fig:1-1}), in view of the lack of non-linearities, the Jacobian and consequently its eigenvalues, are constant throughout its state space. As there are no diagonal elements in the arabesque systems, there can not be any attractor. Moreover, in view of the equality of the coefficients used (all equal to ``1''!), the determinant of $J$ is nil. Consequently, the system of steady state equations is indeterminate, and there is thus no non-degenerate steady state, but rather steady lines (of equations $x=y=z=\ldots$).

As regarding the eigenvalues of the Jacobian matrix (see Table 1), one or two of them are nil depending on the parity (odd or even) of the dimension $n$. This can be understood by the fact that in these systems the characteristic equation comprises only of terms with odd or even exponents depending on whether the dimension is odd or even. The rest of the eigenvalues are pairs of conjugated purely imaginary roots. For sufficiently large dimensions ($5$ or $\geq 7$) two or more of these pairs display non-commensurable values, giving as a result non-commensurable periods: trajectories are then quasi-periodic, since they fail to close on themselves on the tori they generate, and thus cover it densely. The apparently aberrant simplicity (an ellipse, like for the cases $n=3$ and $n=4$) of the trajectories of the $6$-dimensional system can be understood by the fact that the two pairs of conjugated roots have the same values ($\pm \imath\sqrt{3}$), since their characteristic equation is simply given by 
$\lambda^2(\lambda^2+3)^2=0$.

\begin{table}[h]\label{Table1}
\tbl{Jacobian matrix $J$ of the linear arabesque system A0 for different dimensions and its corresponding eigenvalues.}
{\begin{tabular}{cc}\\[-2pt]
\toprule
Dimensionality & Eigenvalues\\[6pt]
\hline\\[-2pt]
3 & $\begin{array}{c}
\pm \imath\sqrt{3}\\
0
\end{array}$
\\\hline\\[-2pt]
4 & $\begin{array}{c}
\pm2\imath\\
0\\
0
\end{array}$
\\\hline\\[-2pt]
5 & $\begin{array}{c}
\pm \imath\sqrt{\frac{1}{2}(5+\sqrt{5})}\\
\pm \imath\sqrt{\frac{5}{2}-\frac{\sqrt{5}}{2}}\\
0
\end{array}$
\\\hline\\[-2pt]
 6 & $\begin{array}{c}
\pm \imath\sqrt{3}\\
\pm \imath\sqrt{3}\\
0\\
0
\end{array}$
\\\hline\\[-2pt]
7 & $\begin{array}{c}
\pm1.94986\imath\\
\pm1.56366\imath\\
\pm0.867767\imath\\
0
\end{array}$
\\\hline\\[-2pt]
8 & $\begin{array}{c}
\pm2\imath\\
\pm \imath\sqrt{2}\\
\pm \imath\sqrt{2}\\
0\\
0
\end{array}$\\[1pt]
\botrule
\end{tabular}}
\end{table}

\begin{center}
\begin{figure}
\center
\includegraphics[width=0.9\textwidth]{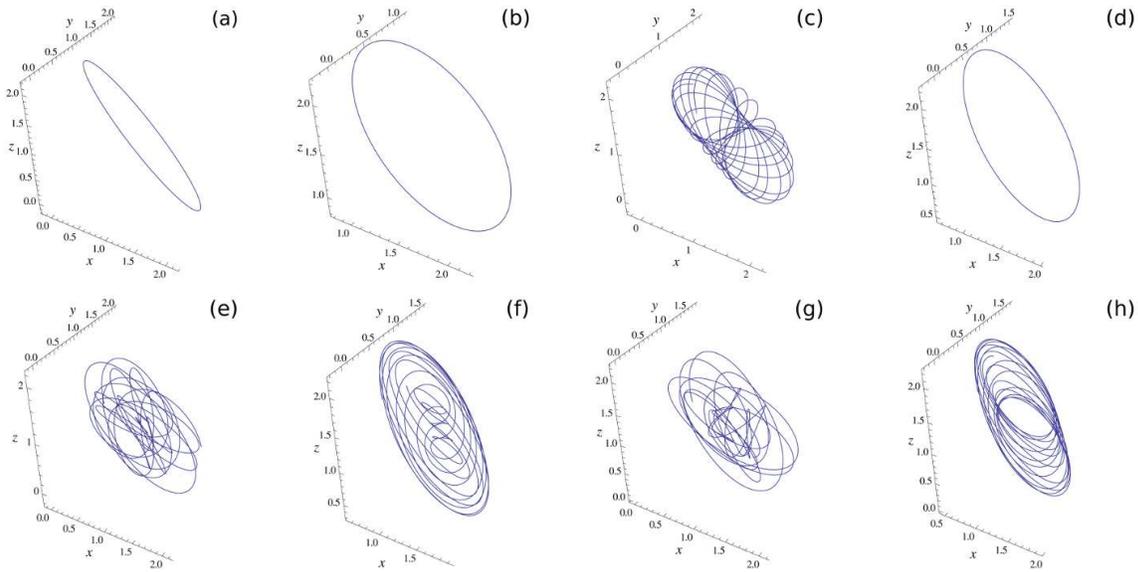}
\caption{3D projections on axes $x,\;y$ and $z$ for 3D to 10D of the linear variant (as in system of Fig. \ref{fig:1-1}). First row, from $n=3$ (panel (a)) to $n=6$ (panel (d)) and second row, from $n=7$ (panel (e)) to $n=10$ (panel (h)).\label{fig:1}}
\end{figure}
\end{center}

A general expression for the eigenvalues $\lambda_{k}$ of the Jacobian matrix of the $n$D linear arabesque system is:
\begin{equation}
\lambda_{k}=2 \imath\sin\Bigl(\frac{2\pi k}{n}\Bigr),k=0,1,\ldots,n-1.\label{eq:eigenvalues_A0_nD}
\end{equation}
This result and Fig. \ref{fig:1} show that linear arabesque systems of odd dimensions display more complex trajectories than the adjacent even-dimensional ones. This can be understood in terms of the presence of an $n$-element negative circuit. Like in the classical case of ellipses around a center in phase space, the size of the tori depends on the initial state but, in contrast to chaotic attractors, they do not display sensitive dependence on initial states. Fig. \ref{fig:1} shows projections of the trajectories (from $n=3$ to $n=10$ dimensions) in the first three axes $x,y,z$ for an initial state $(1,0,2,1,\ldots)$ and with $x_{i}(0)=1,\;i>4$ for higher dimensions. 

\begin{center}
\begin{figure}
\center
\includegraphics[width=0.6\textwidth]{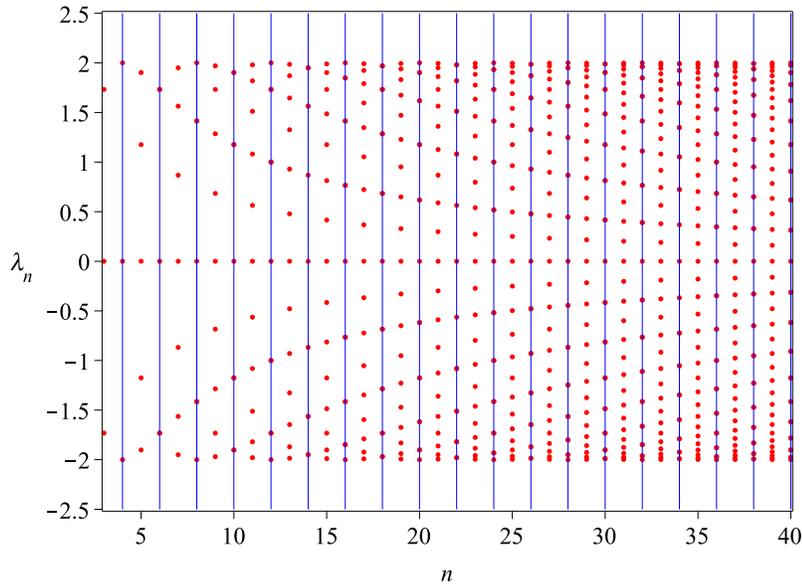}
\caption{The spectra of the eigenvalues $\{\lambda_k\}_n$ for $k=1,\ldots$ and different dimensions $n=3,4,\dots,40$ demonstrate the structural degeneracies due to the symmetries and parity of the corresponding Jacobian matrices.\label{fig:eigenvaluesJ41}}
\end{figure}
\end{center}

The spectra of the Jacobian matrices for increasing dimensions clearly show the same strong degeneracy observed for the non-zero eigenvalues for even dimensions. However, for odd dimensions the degeneracy appears only for the zeroth eigenvalue as in Fig. \ref{fig:eigenvaluesJ41}.

\subsection{One cubic nonlinearity (system A1\_3D)}

The next step in our study is to introduce a nonlinearity that will have no effect on the signs of the circuits of the linear arabesque system of Fig. \ref{fig:1-1}. This is easily realized by rendering cubic the first term $y$ of the first equation $\frac{dx}{dt}$ of this system. We thus obtain for the A1\_3D system:
\begin{eqnarray}
\frac{dx}{dt} & = & y^{3}-z\nonumber \\
\frac{dy}{dt} & = & z-x,\label{eq:A1_3D}\\
\frac{dz}{dt} & = & x-y\nonumber 
\end{eqnarray}
whose Jacobian matrix is:
\begin{equation}
J=\begin{pmatrix}0 & 3y^{2} & -1\\
-1 & 0 & 1\\
1 & -1 & 0
\end{pmatrix}.
\end{equation}
We call this system ``arabesque 1\_3D'' and for short A1\_3D. Consistent with the existence of positive circuit(s), this nonlinearity results in the occurrence of three unstable steady states, as expected from the absence of any attractor. 

The determinant of the Jacobian matrix is no more nil and so the system of steady state equations is no more indeterminate. Moreover, consistent with the simultaneous presence of a nonlinearity and of positive circuit(s),
the system now displays multistationarity meaning that there are three collinear steady states, all unstable as expected by the absence of any attractor.

\begin{table}[h]\label{Table2}
\tbl{Steady states of system \eqref{eq:A1_3D} and their corresponding approximate eigenvalues.}
{\begin{tabular}{c c}\\[-2pt]
\toprule
Steady state & Eigenvalues of steady state\\[6pt]
\hline\\[-2pt]
$(-1,-1,-1)$ & $-0.3882\ldots,\:0.1941\ldots\pm2.2612\ldots\imath$
\\\hline\\[-2pt]
$(0,0,0)$ & $0.4533\ldots,\:-0.2266\ldots\pm1.4677\ldots\imath$
\\\hline\\[-2pt]
$(1,1,1)$ & $-0.3882\ldots,\:0.1941\ldots\pm2.2612\ldots\imath$\\[1pt]
\botrule
\end{tabular}}
\end{table}

As we have already pointed out, the state-space of system \eqref{eq:A1_3D} can be partitioned into domains according to the nature of the eigenvalues of its Jacobian matrix (see Sec. \ref{sub:=00201CFrontiers=00201D:-Partition-of}). We recall that frontier $F_{1}$ is the locus of points at which one or more eigenvalues of $J$ are nil, $F_{2}$ is the locus at which there is a pair of purely imaginary complex conjugated eigenvalues of $J$ and frontier $F_{4}$ is the locus of points where two or more eigenvalues of $J$ are equal. For system \eqref{eq:A1_3D}, the analytic descriptions of these frontiers take the form:
\begin{equation}
\begin{array}{c}
F_{1}\\
F_{2}\\
F_{4}
\end{array}\begin{array}{c}
:\\
:\\
:
\end{array}\begin{array}{c}
-1+3y^{2}=0\\
-1+3y^{2}=0\\
27(1-3y^{2})^{2}+4(2+3y^{2})^{3}=0
\end{array}.
\end{equation}
We observe that $F_{1}$ and $F_{2}$ coincide. These frontiers define two planes given by $y=\pm\frac{1}{\sqrt{3}}$. Between them, the signs of the eigenvalues are of the type $\{+/--\}$ and outside them, they are of the type $\{-/++\}$. There is thus at the level of these frontiers, a shift from $+$ to $-$ of a real eigenvalue (i.e. the effect of frontier $F_{1}$) and a shift from $--$ to $++$ of the real part of a pair of complex conjugated eigenvalues (i.e. the effect of frontier $F_{2}$). There is no $F_{4}$ frontier since function $F_{4}$ is strictly positive for all real $y$, in agreement with the fact that the eigenvalues remain complex everywhere in the state-space of system \eqref{eq:A1_3D}.

Strictly speaking, what the frontiers tell us is that the nature of any steady state that might be present in a domain of the partition is determined by the sign pattern which is characteristic of this domain. However, trajectories are not ``prisoners'' of a frontier. For example, in dynamical systems such as for example the well-known R\"{o}ssler system, trajectories alternatively evolve around two unstable steady states that belong to distinct domains.

In order to get a picture of the dynamics of the nonlinear arabesque system \eqref{eq:A1_3D}, we have performed a series of numerical integrations of a large ensemble of initial states and found that its trajectories are confined in two compact domains which are located symmetrically about the origin which is the center of symmetry. Due to its special symmetry properties, to each trajectory starting from an initial state ${(x_{0},y_{0},z_{0})}$ corresponds an ``opposite'' trajectory starting from the opposite initial state ${(-x_{0},-y_{0},-z_{0})}$, which is symmetric with respect to the center of symmetry (i.e. to the origin). These point-symmetric domains lie between the two ``outer'' steady states ${(1,1,1)}$ and ${(-1,-1,-1)}$.

\begin{figure}
\center\includegraphics[width=0.3\textwidth]{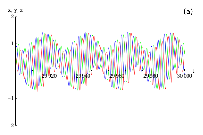}
\includegraphics[width=0.3\textwidth]{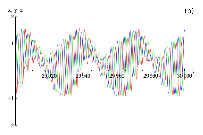}
\includegraphics[width=0.3\textwidth]{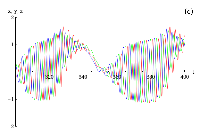}\\
\includegraphics[width=0.3\textwidth]{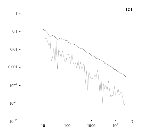}
\includegraphics[width=0.3\textwidth]{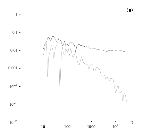}
\includegraphics[width=0.3\textwidth]{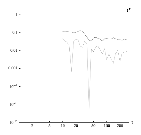}
\caption{Time evolution of variables $x$, $y$ and $z$ of system \eqref{eq:A1_3D} superimposed with different colors, for the initial state: (a) ${(0.7,0,0)}$ which is located inside the compact domain, (b) ${(-0.2,0.6,0)}$ which lies at the periphery of the compact domain and (c) ${(0.3,0,0)}$ which is located outside the compact domain, between $t=29900$ and $t=30000$ for panels (a) and (b), and between $t=300$ and $t=400$ for panel (c). Evolution of the maximum Lyapunov exponent $\sigma_{1}$ (black) and second maximum $\sigma_{2}$ (gray) for: (d) the initial state of panel (a), (e) the initial state of panel (b) and (f) the initial state of panel (c). Note that all axes of the last three panels are in log-log scale except the horizontal of the last panel.\label{fig:2}}
\end{figure}

\begin{figure}
\center
\includegraphics[width=0.3\textwidth]{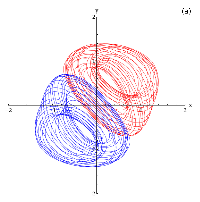}
\includegraphics[width=0.3\textwidth]{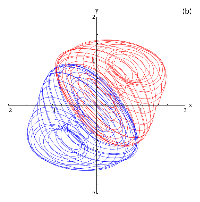}
\includegraphics[width=0.3\textwidth]{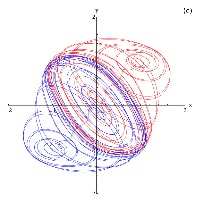}
\caption{$xy$ plots of trajectories of initial states ${(0.7,0,0)}$ in panel (a), ${(-0.2,0.6,0)}$ in panel (b) and ${(0.3,0,0)}$ in panel (c) (in red) and of their ``opposites'' ${(-0.7,0,0)}$, ${(0.2,-0.6,0)}$ and ${(-0.3,0,0)}$ (in blue).\label{fig:3}}
\end{figure}

We show typical time evolutions of three such characteristic trajectories of system \eqref{eq:A1_3D} in Fig. \ref{fig:2}. They are seen to originate, respectively, from initial states located (a) inside (i.e. $(x,y,z)=(0.7,0,0)$), (b) at the periphery (i.e. $(x,y,z)=(-0.2,0.6,0)$) and (c) outside (i.e. $(x,y,z)=(0.3,0,0)$) of the compact domains mentioned above. In contrast with trajectories (a) and (b), the third one escapes to infinity at about $t=500$, and that is why it is plotted only from $t=300$ to $t=400$. In panels (d) to (f) of Fig. \ref{fig:2}, we present the time evolution of the corresponding Lyapunov exponents $\sigma_{1}$ and $\sigma_{2}$ of cases (a) to (c) respectively. We deduce that the trajectory which is initially located inside the compact domain (see panel (a) and (d) of Fig. \ref{fig:2}) is ordered at least up to $t=3\cdot10^{4}$ since its maximum Lyapunov exponent $\sigma_{1}$ decreases towards zero following a power law $\propto\frac{1}{t}$, the trajectory which is initially located 
at the periphery of the compact domain (see panel (b) and (e) of Fig. \ref{fig:2}) shows a chaotic behavior since its $\sigma_{1}$ converges to a positive value at least up to $t=3\cdot10^{4}$ and finally the trajectory which is initially located outside the compact domain (see panel (c) and (f) of Fig. \ref{fig:2}) presents a transient chaotic behavior before it escapes to infinity at about $t=400$. The $xy$ plots of the trajectories that correspond to the plots of panels (a), (b) and (c) (in red) of Fig. \ref{fig:2} and their ``opposite'' (in blue) are shown in panels (a), (b) and (c) of Fig. \ref{fig:3}.

\begin{figure}
\center\includegraphics[width=0.5\textwidth]{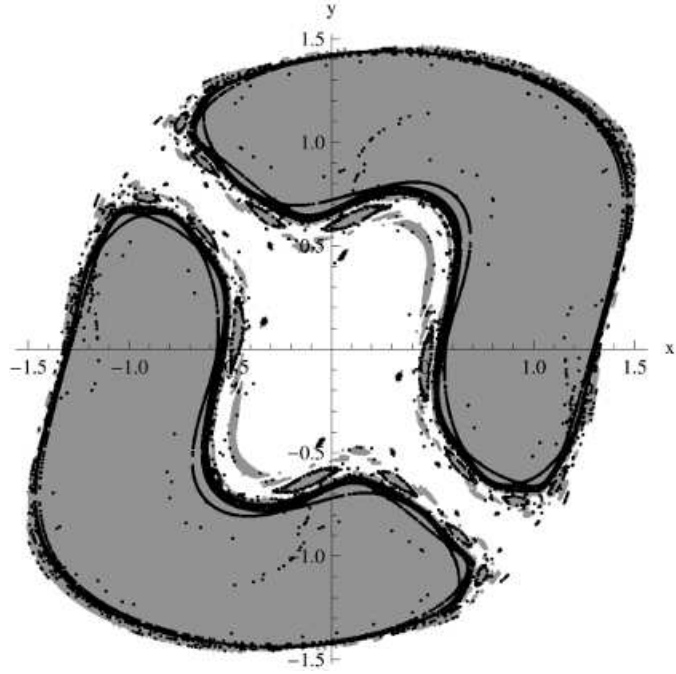}
\caption{Cross section of points $(x,y)$ with $z=0$ of the state-space of the nonlinear arabesque system \eqref{eq:A1_3D} using the SALI method. Each point represents an initial state integrated up to $t_{f}=10^{6}$. Gray points represent regular (quasi-periodic or periodic) initial states with $\mathrm{SALI}(t_{f})\geq10^{-8}$, black chaotic initial states with $\mathrm{SALI}(t_{f})<10^{-8}$ while white corresponds to initial states that escape to infinity before $t_{f}$.\label{fig:6}}
\end{figure}

In order to have a picture of the dynamics of the conservative arabesque system \eqref{eq:A1_3D}, we iterated in Fig. \ref{fig:6} $10^{6}$ initial states located on a rectangular equally spaced grid for $(x,y)\in[-1.5,1.5]\times[-1.5,1.5]$ with $z=0$ up to final integration time $t_{f}=10^{6}$. We computed their $\mathrm{SALI}(t)$ values following the methodology described in Subsec. \ref{sub:The-SALI-method} and focused on those that do not escape to infinity earlier than $t_{f}$. We categorized the non-escaped trajectories as follows: if $\mathrm{SALI}(t_{f})\geq10^{-8}$ then we characterize the initial state as regular (quasi-periodic or periodic) and color it gray in the plot of Fig. \eqref{fig:6}, otherwise (i.e. if $\mathrm{SALI}(t_{f})<10^{-8}$) as chaotic and color it black. White color corresponds to initial states that escape to infinity before $t_{f}$. In order to apply this kind of binary filtration, we used as a chaos threshold the value $10^{-8}$ as it has been used 
successfully in previous publications \cite{Skokos2001,Skokosetal2004} as a good one for the efficient discrimination between regularity and chaos in Hamiltonian systems, symplectic maps and galactic potentials \cite{Skokos2001,Skokosetal2004,CapuzzoDolcettaetal2007,Manosetal2008,Manosetal2011}.

In Fig. \ref{fig:6}, we can see that there are two point-symmetric ``kidney''-shaped big regions of regular motion each one bounded by its own point-symmetric thin chaotic layers. These chaotic layers cover the region of regular initial states and create complex structures of chains of stability islands of small size and of high period. Outside the two main big regular regions there are smaller stability islands of higher period which are surrounded by their own even thinner chaotic layers. It is interesting to note that trajectories starting initially from the interior of the regular region (i.e. inside the ``kidney''-shaped region of Fig. \ref{fig:6}) are not able to penetrate the surrounding thin chaotic layer and subsequently escape to infinity later on, and so this property permit us to refer to the ``kidney''-shaped regions as compact regions.

\begin{figure}
\center\includegraphics[width=0.5\textwidth]{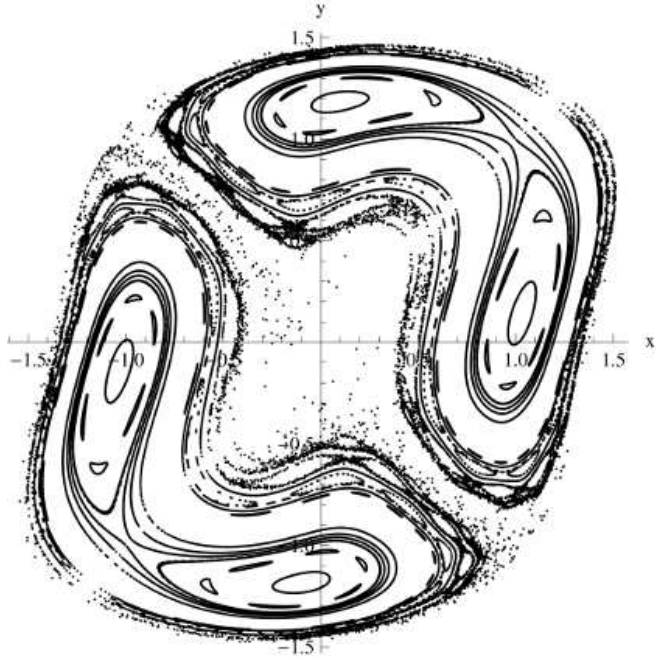}
\caption{PSS of points $(x,y)$ satisfying $z=0$ of the state-space of system \eqref{eq:A1_3D} as a juxtaposition for $z\geq0$ and $z\leq0$ due to its point-symmetric property.\label{fig:7}}
\end{figure}

In Fig. \ref{fig:7} we present the corresponding PSS of points $(x,y)$ of system \eqref{eq:A1_3D} for $z=0$. In particular, we see the special state-space structures (i.e. foliated tori) of the two main regular gray point-symmetric regions of Fig. \ref{fig:6}. Interestingly enough, inside each ``kidney''-shaped region, there are two main elliptic points (the centers of the big elliptic regions) separated by a hyperbolic one. Around these elliptic points we find again chains of stability islands of high period and of relatively big size depicted as smooth elliptic closed curves. This special structure around the elliptic points is separated by the separatrix coming out of the unstable hyperbolic point. As we move away from the two main elliptic points towards the origin (point of symmetry), we can see similar in nature but even more complex 
thinner structures of chains of stability islands of smaller size and of higher period. This is so until we reach the complex ``border'' of the ``kidney''-shaped region, the so-called ``edge of chaos'' \cite{Bountisetal2012}, outside of which there is a narrow chaotic region filled by randomly scattered points and initial states that escape to infinity.

\begin{figure}
\center
\includegraphics[width=0.4\textwidth]{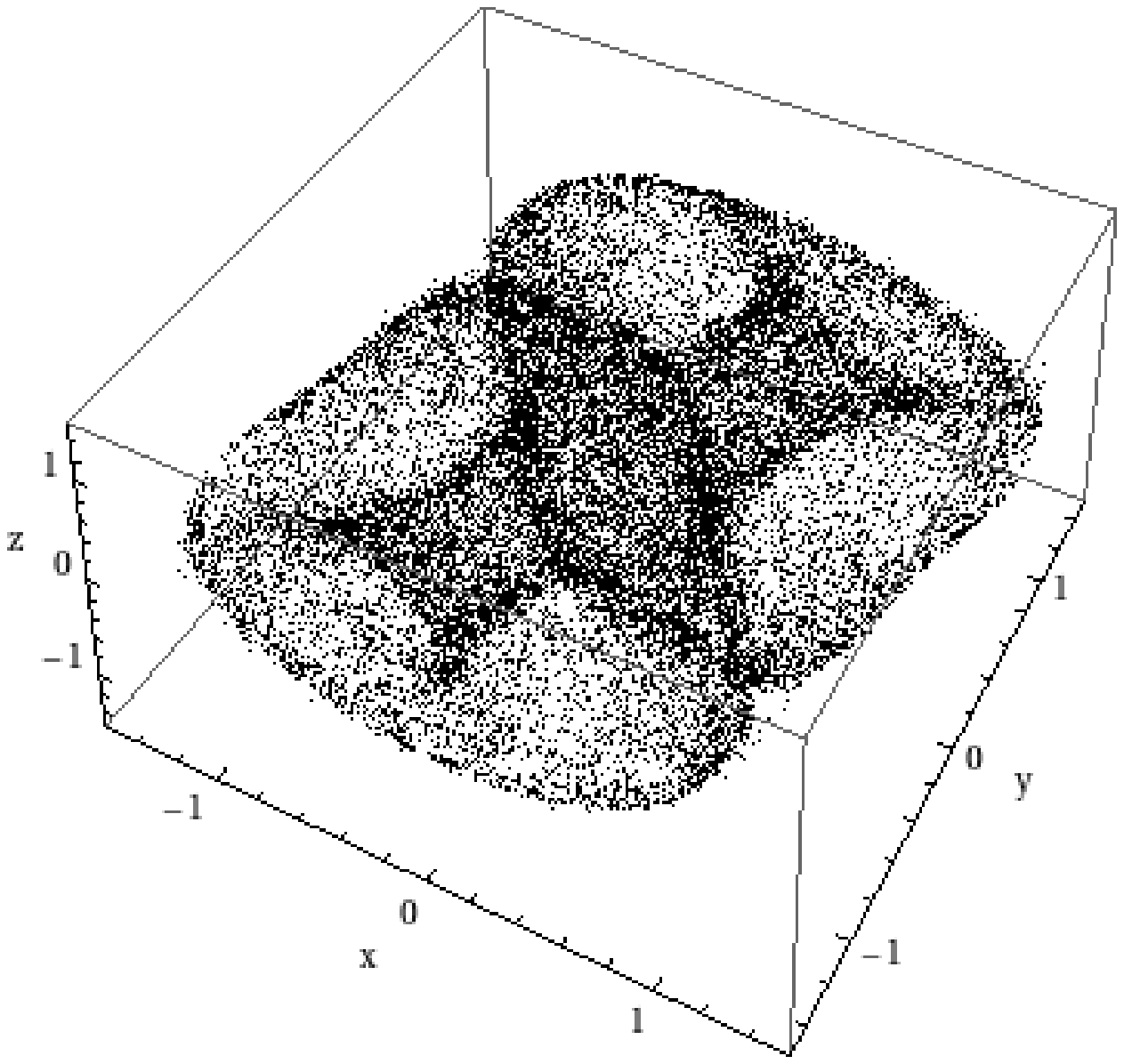}
\includegraphics[width=0.4\textwidth]{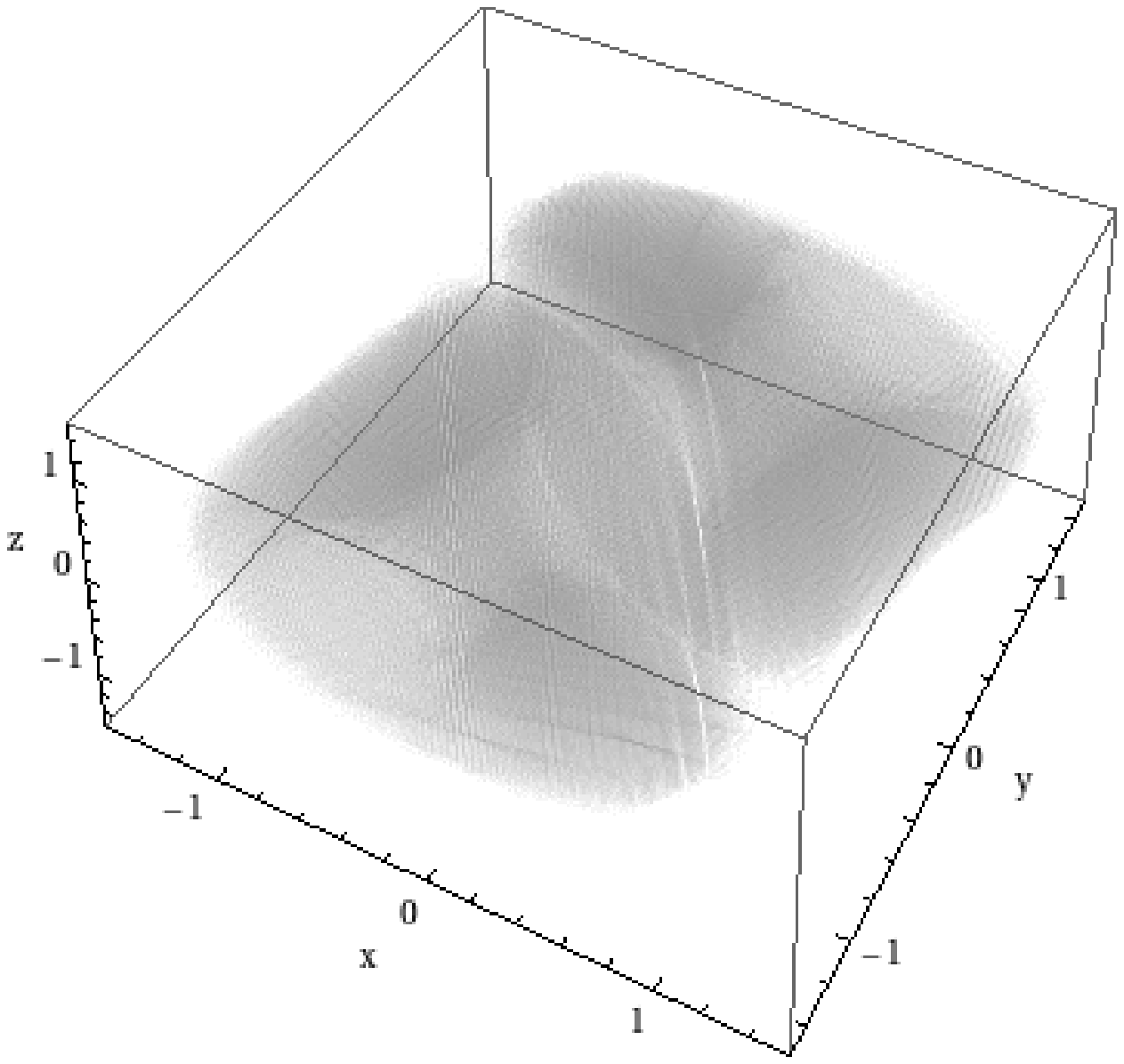}
\caption{Volume of chaotic initial states of the nonlinear arabesque system \eqref{eq:A1_3D} in panel (a). We have located them keeping only those that give $\mathrm{SALI}(t_{f})<10^{-8}$. Volume of regular initial states of the same arabesque system in panel (b). We found them keeping only those that give $\mathrm{SALI}(t_{f})\geq10^{-8}$. In this plot we considered as final integration time $t_{f}=10^{5}$.\label{fig:8}}
\end{figure}

To pursue this finding further, we have first approximated the layer of all initial states $(x,y,z)$ of system \eqref{eq:A1_3D} which lead to chaotic behavior. This was achieved by considering initial states on an equally spaced cubic grid $(x,y,z)\in[-1.5,1.5]\times[-1.5,1.5]\times[-1.5,1.5]$, keeping only those whose trajectories are characterized by a value of $\mathrm{SALI}(t_{f})<10^{-8}$ (see Fig. \ref{fig:8} (a)). Then, we computed the fractal dimension \cite{Sarrailleetal1994} of this set of points that approximates the thin chaotic layer and found that it is approximately $2.4\pm0.025$. Such a value signifies a fractal object clearly located between 2 and 3 dimensions. By doing the same for one of the two symmetric main volumes of regular initial states (see Fig. \ref{fig:8} (b)) we have found that its fractal dimension is numerically very close to a three dimensional object since FD3 program reports $2.98\pm0.3$ (see Fig. \ref{fig:8} (b)).

\begin{figure}
\center
\includegraphics[width=0.4\textwidth]{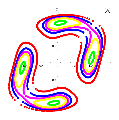}
\includegraphics[width=0.4\textwidth]{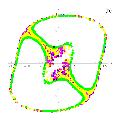}
\caption{PSS on the plane $(x,y)$ with $z=0$ of five pairs of ``opposite'' trajectories (from $t=0$ up to $t=1000$) of system \eqref{eq:A1_3D}, starting from inside the ``kidneys'' in panel (a). PSS on the plane $(x,y)$ with $z=0$ of five pairs of ``opposite'' trajectories starting from the edge of the ``kidney''-shaped regions in panel (b). Each color corresponds to points of the same trajectory.\label{fig:9}}
\end{figure}

The fact that most of the trajectories starting from the interior of the point-symmetric ``kidney''-shaped regions are quasi-periodic and most of those on the edge of the ``kidney''-shaped regions chaotic, implies that these trajectories are confined in the domains they come from and consequently do not ``merge'' with each other. One can thus expect that trajectories starting from inside the ``kidney''-shaped regions should fill these two point-symmetric volumes and consequently these volumes should be compact. This could be checked by computing the corresponding PSS, that is a selection of points of trajectories (one color for each individual trajectory) that lie for example on the plane $(x,y)$ with $z=0$ (see Fig. \ref{fig:9} (a) and (b)). The two panels of Fig. \ref{fig:9} give an idea of the complex structure of the core and of the edge of the ``kidney''-shaped regions. The left panel shows the points of quasi-periodic trajectories, located in the core of the ``kidney''-shaped regions while the right 
panel those of chaotic trajectories, located in the edge.

Systems of type A1\_$n$D with $n>3$ have been studied in less detail, however. For odd dimensions, they behave essentially, {\em mutatis mutandis}, like A1\_3D, including the number of steady states, which remain three, consistent with the degree of the unique nonlinearity. Systems A1\_$n$D with even dimensions have two nil eigenvalues, like the linear systems. The determinant of the Jacobian matrix is nil, and consequently the steady state equations of the system are indeterminate. There are in fact three steady lines in 4D (i.e a line $x=z$ in each of the planes $y,u={-1,1,1}$) and in 6D there are three lines $x=z=u$ etc.. This multi modality is understandable in view of the simultaneous presence of a nonlinearity and of positive circuits.

It is of interest to remark that, in these even-dimensional systems, it was possible to eliminate the positive circuits simply be permuting the signs of the elements of any of the 2-element circuits. As the determinant of $J$ is no more nil, the system is no more indeterminate, and there is a single steady state, in view of the absence of positive circuits.

\subsection{Systems with $n$ cubic nonlinearities (A2\_3D \& A2\_4D)}
\subsubsection{System A2\_3D}

\begin{figure}[!ht]
\center\includegraphics[width=0.5\textwidth]{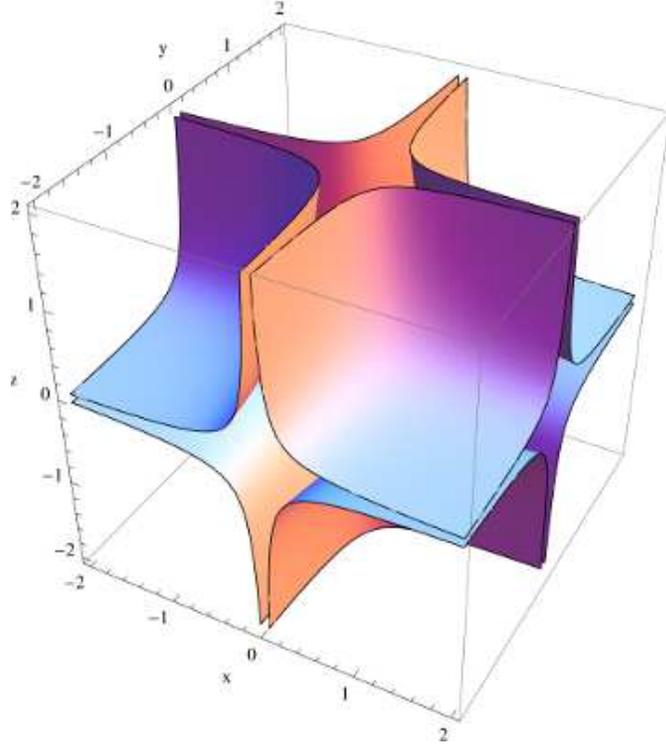}
\caption{Plot of frontiers $F_{1}$ and $F_{2}$ of system \eqref{eq:A2_3D} which coincide. In the region confined by the two frontiers, the real eigenvalue is negative and the real parts of the complex eigenvalues are positive $\{-/++\}$ like it is for the steady state $(0,0,0)$. Outside the eight hyperbolic surfaces, the real eigenvalue is positive and the real parts of the complex ones are negative $\{+/--\}$, like at the steady states $(1,1,1)$ and $(-1,-1,-1)$.\label{fig:12}}
\end{figure}

In this subsection, instead of inserting an extra nonlinearity to the second variable in the first equation, as we have done in the previous section, we apply a cubic nonlinearity to each positive term of the arabesque system \eqref{eq:A1_3D}. This slightly alters the simplicity of the system, but, on the other hand, increases the number of its symmetries. In 3D, the equations read:
\begin{eqnarray}
\frac{dx}{dt} & = & y^{3}-z\nonumber\\
\frac{dy}{dt} & = & z^{3}-x,\label{eq:A2_3D}\\
\frac{dz}{dt} & = & x^{3}-y\nonumber 
\end{eqnarray}
and its Jacobian matrix is:
\begin{equation}
J=\begin{pmatrix}0 & 3y^{2} & -1\\
-1 & 0 & 3z^{2}\\
3x^{2} & -1 & 0
\end{pmatrix}.
\end{equation}
The presence of three nonlinearities in \eqref{eq:A2_3D} does not increase substantially the complexity of its behavior and similarly to system \eqref{eq:A1_3D}, there are still three steady states $(-1,-1,-1)$, $(0,0,0)$, $(1,1,1)$ with its trajectories being confined again inside two ``kidney''-shaped volumes. The analytical expression of the new frontiers $F_{1}$, $F_{2}$ and $F_{4}$ are:
\begin{equation}
\begin{array}{c}
F_{1}\\
F_{2}\\
F_{4}
\end{array}\begin{array}{c}
:\\
:\\
:
\end{array}\begin{array}{c}
-1+27x^{2}y^{2}z^{2}=0\\
-1+27x^{2}y^{2}z^{2}=0\\
4(3x^{2}+3y^{2}+3z^{2})^{3}+27(1-27x^{2}y^{2}z^{2})^{2}=0
\end{array}.
\end{equation}
Frontiers $F_{1}=0$ and $F_{2}=0$ are no more simple planes, but eight hyperbolic surfaces (see Fig. \ref{fig:12}). However, like system \eqref{eq:A1_3D}, there are only two types of eigenvalues of $J$: either $\{+/--\}$ or $\{-/++\}$. This is why frontiers $F_{1}=0$ and $F_{2}=0$ coincide and there is no frontier $F_{4}=0$ as it is strictly positive for all $x,y$ and $z$. These similarities between systems \eqref{eq:A1_3D} and \eqref{eq:A2_3D} are not surprising, as far as one realizes that the signs of all circuits involved are of the same type in the two systems (see Fig. \ref{fig:12}).

Table 3 summarizes the eigenvalues of system \eqref{eq:A2_3D} at its steady states and Fig. \ref{fig:12} shows frontiers $F_{1}$ and $F_{2}$ which coincide.

\begin{table}[!ht]\label{Table3}
\tbl{Steady states of system \eqref{eq:A2_3D} and their corresponding eigenvalues.}
{\begin{tabular}{c c}\\[-2pt]
\toprule
Steady State & Eigenvalues of steady state\\[6pt]
\hline\\[-2pt]
$(-1,-1,-1)$ & $2,-1\pm2\sqrt{3}\imath$
\\\hline\\[-2pt]
$(0,0,0)$ & $-1,0.5\pm\frac{\sqrt{3}}{2}\imath$
\\[2pt]\hline\\[-2pt]
$(1,1,1)$ & $2,-1\pm2\sqrt{3}\imath$\\[1pt]
\botrule
\end{tabular}}
\end{table}

\subsubsection{System A2\_4D}

The 4D case of the nonlinear arabesque system which we call for short A2\_4D is given by:
\begin{equation}
\begin{array}{c}
\frac{dx}{dt}\\
\frac{dy}{dt}\\
\frac{dz}{dt}\\
\frac{du}{dt}
\end{array}\begin{array}{c}
=\\
=\\
=\\
=
\end{array}\begin{array}{c}
y^{3}-u\\
z^{3}-x\\
u^{3}-y\\
x^{3}-z
\end{array}\label{eq:A2_4D}
\end{equation}
and its Jacobian matrix by:
\begin{equation}
J=\begin{pmatrix}0 & 3y^{2} & 0 & -1\\
-1 & 0 & 3z^{2} & 0\\
0 & -1 & 0 & 3u^{2}\\
3x^{2} & 0 & -1 & 0
\end{pmatrix}.
\end{equation}
Here and in fact, in all A2 systems with an even dimension, both $n$-element circuits are positive, and, in agreement with this, the number of steady states increases from three in A2\_3D to nine in A2\_4D. Table 4 shows these steady states and their corresponding eigenvalues. As one can see, the steady states can be classified into three groups regarding the signs (and in fact the values) of the eigenvalues of $J$. Steady state $(0,0,0)$ has two real, namely $\pm1$, and two purely imaginary, namely $\pm\imath$, eigenvalues. The four ``external'' steady states (i.e. those whose coordinates are all $\pm1$) also have two real, namely $\pm$2 and two purely imaginary, namely $\pm4 \imath$, eigenvalues. The ``intermediate'' steady states (i.e. those whose coordinates are $\pm1$ or $0$) display two pairs of conjugate purely imaginary eigenvalues, namely $\pm2\imath$ and $\pm\sqrt{2} \imath$.

\begin{table}[ht!]\label{Table4}
\tbl{Steady states of system \eqref{eq:A2_4D} and their corresponding eigenvalues.}
{\begin{tabular}{c c}\\[-2pt]
\toprule
Steady state & Eigenvalues of steady state\\[6pt]
\hline\\[-2pt]
$(-1,-1,-1,-1)$ & $\pm4\imath,\pm2$
\\\hline\\[-2pt]
$(-1,0,-1,0)$ & $\pm2\imath,\pm\sqrt{2}\imath$
\\\hline\\[-2pt]
$(-1,1,-1,1)$ & $\pm4\imath,\pm2$
\\\hline\\[-2pt]
$(0,-1,0,-1)$ & $\pm2\imath,\pm\sqrt{2}\imath$
\\\hline\\[-2pt]
$(0,0,0,0)$ & $\pm\imath,\pm1$
\\\hline\\[-2pt]
$(0,1,0,1)$ & $\pm2\imath,\pm\sqrt{2}\imath$
\\\hline\\[-2pt]
$(1,-1,1,-1)$ & $\pm4\imath,\pm2$
\\\hline\\[-2pt]
$(1,0,1,0)$ & $\pm2\imath,\pm\sqrt{2}\imath$
\\\hline\\[-2pt]
$(1,1,1,1)$ & $\pm4\imath,\pm2$\\[1pt]
\botrule
\end{tabular}}
\end{table}

The analytical expression of the frontiers of system \eqref{eq:A2_4D} are:
\begin{equation}
\begin{array}{c}
F_{1}\\
F_{2}\\
F_{4}
\end{array}\begin{array}{c}
:\\
:\\
:
\end{array}\begin{array}{c}
-1+9u^{2}y^{2}+9x^{2}z^{2}-81x^{2}y^{2}z^{2}u^{2}z^{2}=0\\
\varnothing\\
-1+9u^{2}y^{2}+9x^{2}z^{2}-81x^{2}y^{2}z^{2}u^{2}=0
\end{array}.
\end{equation}
As one can see, $F_{2}$ is nil everywhere (i.e. there is no $F_{2}$ frontier), and frontiers $F_{1}$ and $F_{4}$ coincide. This describes the fact that there are complex eigenvalues everywhere, and that two adjacent domains of the state-space of system \eqref{eq:A2_4D} differ. At the level of the double $F_{1}$-$F_{4}$ frontier, one pair of real eigenvalues vanishes and beyond this limit it is replaced by a pair of conjugated, purely imaginary eigenvalues.

The steady states can be visualized, for example, by sections following the three planes $(y=-1,\;u=-1),\;(y=0,\;u=0)$ and $(y=1,\;u=1)$. In either of these planes, there are three steady states whose $x$ and $z$ coordinates are $(-1,-1),\;(0,0)$ and $(1,1)$.

\begin{figure}[!ht]
\center
\includegraphics[width=0.4\textwidth]{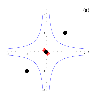}
\includegraphics[width=0.4\textwidth]{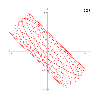}
\caption{Section showing the plane ($y=-1,\;u=-1$) with its three steady states ${(-1,-1,-1,-1)}$, ${(0,-1,0,1)}$, ${(1,-1,1,-1)}$ as black points in panel (a). The blue curves are the section of frontiers $F_{1}$ and $F_{4}$. From initial states in the vicinity of state ${(0,-1,0,-1)}$ of panel (a), one gets closed trajectories, one of them shown in panel (b).\label{fig:13}}
\end{figure}

\begin{figure}
\center
\includegraphics[width=0.4\textwidth]{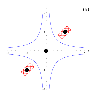}
\includegraphics[width=0.4\textwidth]{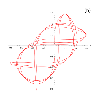}
\caption{Section showing the plane ($y=0,\;u=0$) with its three steady states $(-1,0,-1,0)$, $(0,0,0,0)$, $(1,0,1,0)$ as black points in panel (a). The blue curves are the section of frontiers $F_{1}$ and $F_{4}$. From initial states in the vicinity of states ${(-1,0,-1,0)}$ and ${(1,0,1,0)}$ of panel (a), one gets regular trajectories, one of them shown in panel (b).}\label{fig:14}
\end{figure}

Figs. \ref{fig:13} and \ref{fig:14} show respectively in plane ($y=-1$, $u=-1$) and ($y=0$, $u=0$), (a) the frontiers and steady states and (b) trajectories from initial states close to steady states (these trajectories are also seen in (a) at the level of the relevant steady states). The situation in plane ($y=1$, $u=1$) has not been represented since it is simply symmetrical to that of plane ($y=-1$, $u=-1$). One sees that compact trajectories (in red) are generated from initial states close to the four steady states whose coordinates include both 0 and $\pm 1$ (thus, the four steady states whose eigenvalues are two pairs of conjugated purely imaginary eigenvalues).

\section{Conclusions and discussion on open problems}

In this paper we have studied a family of dynamical systems that we call ``arabesques'', whose Jacobian matrices possess closed chains of 2-element negative circuits. In view of the absence of diagonal terms in their Jacobian matrices, all these dynamical systems are conservative and consequently, they do not posses attractors. First, we have analyzed the linear variant A0 of this family and then, we have inserted a single, cubic nonlinearity in one of its equations so that the signs of its circuits are not affected. We have analyzed the new nonlinear system in some detail and found that it displays a complex mixed set of quasi-periodic and chaotic trajectories. Next, we have inserted $n$ cubic nonlinearities, one per equation, and generated in particular the A2\_3D arabesque system. We have shown that it has three steady states in agreement with the order of the system. In fact, it behaves essentially as A1\_3D does, since they share the same circuit signs.

We then passed to the system A2\_4D, that has two positive $4$-circuits and discussed about its dynamics in terms of its Jacobian circuits and frontiers. Finally, we have also investigated and compared the complex dynamics of this family of dynamical systems in terms of their symmetries. We conclude here by mentioning some open problems which are related to our work and we think they deserve to be addressed in the future.

The first one is the role of functions $C$ and $H_{a+1}$, for which we see that an ``energy'' dissipation or conservation condition is not merely due to the presence of a nonlinearity, but depends solely on the symmetry breaking between the nonlinearity of the variables. In view of our expertise in Hamiltonian dynamics via SALI, GALI, Lyapunov Exponents and fractal dimensions, we propose as a future line of research to investigate the role of circuits of Jacobian matrices in classical and quantum mechanical multidimensional Hamiltonians (like Bose-Einstein condensate versus Fermi-Pasta-Ulam lattices) \cite{Antonopoulosetal2006A,Antonopoulosetal2006B,Skokosetal2008} and their connection to the time evolution of the variational equations of the previously mentioned chaos detection methods as well as to study the influence of a small additional potential term (e.g. due to an external field) \cite{Demongeotetal2007}.

The second is related to the case of Boolean tangential negative circuits of any length where we have observed the occurrence of limit-cycles of long period, analogous to those of the continuous case \cite{Demongeotetal2012A}. These limit cycles conserve both the discrete kinetic energy and the global frustration function of the network, in the case of a Hopfield dynamics \cite{Demongeotetal2012B}. After this observation, an interesting further investigation suggests by itself: The study of common features observed in both Boolean and continuous cascades of negative loops frequently observed in genetic and epigenetic regulations \cite{Cinquinetal2002} and try to extend the present results to circuits intersecting in two or more points, showing a smaller number of invariant trajectories than in the tangential case \cite{Demongeotetal2011}.

The last one is that symmetries found in the Jacobian matrix affect the eigenvalues and frontiers and a comparison with symmetries observed in known chaotic Chua's circuits and symplectic maps could be further envisioned in order to check if their influence on the eigenspectrum is of the same nature.

\section*{Acknowledgements}

We wish to dedicate this paper to Prof. Otto R\"{o}ssler as the continuing source of inspiration and kind guide that he is. R. T. wishes to thank his grand-daughter Lucille Thomas as she is at the origin of this paper due to her critical and observant reading of \cite{Thomasetal2006}, which revived his interest in the arabesque systems. He also thanks Prof. G. Halbout of the University of Montpellier for useful comments about the concept of ``discriminant''. Ch. A. was supported by a grant from G.S.R.T., Greek Ministry of Education, for the project ``Complex Matter'', awarded under the auspices of the ERA Complexity-NET and by the EPSRC grant: ``EPSRC EP/I032606/1''. V. B. acknowledges support by the European Space Agency under Contract No. ESA AO-2004-070, (No. C90241 and C90238) and the Prodex program of the Belgian Government. This research has been co-financed by the European Union (European Social Fund - ESF) and Greek national funds through the Operational Program ``Education and Lifelong Learning'' of 
the National Strategic Reference Framework (NSRF) - Research Funding Program: THALES - Investing in knowledge society through the European Social Fund.

\bibliographystyle{ws-ijbc}

\begin{thebibliography}{9}

\bibitem[{{Antonopoulos et al}(2006A)}] {Antonopoulosetal2006A} {Antonopoulos, Ch., Bountis, T. \& Skokos, Ch.} [2006A] ``Chaotic dynamics of $N$-degree of freedom Hamiltonian systems,'' {\it Int. J. Bif. Chaos}, {\bf 16}, 6, 1777--1793.

\bibitem[{{Antonopoulos et al}(2006B)}] {Antonopoulosetal2006B} {Antonopoulos, Ch. \& Bountis, T.} [2006B] ``Stability of simple periodic orbits and chaos in a Fermi-Pasta-Ulam lattice,'' {\it Phys. Rev. E}, {\bf 73}, 056206.

\bibitem[{{Antonopoulos et al}(2006C)}] {Antonopoulosetal2006C} {Antonopoulos, Ch. \& Bountis, T.} [2006C] ``Detecting order and chaos by the Linear Dependence Index (LDI) method,'' {\it ROMAI J.}, {\bf 2}, 2, 1--13.

\bibitem[{{Aracena et al}(2004A)}] {Aracenaetal2004A} {Aracena, J., Demongeot, J. \& Goles, E.} [2004A] ``On limit cycles of monotone functions with symmetric connection graph,'' {\it Theor. Comput. Sci.}, {\bf 322}, 2, 237--244.

\bibitem[{{Aracena et al}(2004B)}] {Aracenaetal2004B} {Aracena, J., Demongeot, J. \& Goles, E.} [2004B] ``Positive and negative circuits in discrete neural networks,'' {\it IEEE Trans. Neural Networks}, {\bf 15}, 1, 77--83.

\bibitem[{{Benettin et al}(1976)}] {Benettinetal1976} {Benettin, G., Galgani, L. \& Strelcyn, J-M.} [1976] ``Kolmogorov entropy and numerical experiments,'' {\it Phys. Rev. A}, {\bf 14}, 2338--2345.

\bibitem[{{Benettin et al}(1980A)}] {Benettinetal1980A} {Benettin, G., Galgani, L., Giorgilli, A. \& Strelcyn, J-M.} [1980A] ``Lyapunov characteristic exponents for smooth dynamical systems and for Hamiltonian systems: A method for computing all of them. Part 1: Theory,'' {\it Meccanica}, {\bf 15}, 9--20.

\bibitem[{{Benettin et al}(1980B)}] {Benettinetal1980B} {Benettin, G., Galgani, L., Giorgilli, A. \& Strelcyn, J-M.} [1980B] ``Lyapunov characteristic exponents for smooth dynamical systems and for Hamiltonian systems: A method for computing all of them. Part 2: Numerical application,'' {\it Meccanica}, {\bf 15}, 21--30.

\bibitem[{{Bernard}(1943)}] {Bernard1943} {Bernard, C.} [1943] {\it Introduction \`{a} la M\'{e}decine Exp\'{e}rimentale} (Gibert).

\bibitem[{{Bohn}(1961)}] {Bohn1961} {Bohn, E.} [1961] ``Stability margins and steady-state oscillations of on-off feedback systems,'' {\it IRE Trans. on Circuit Theory}, {\bf 8}, 127--130.

\bibitem[{{Bountis \& Skokos}(2012)}] {Bountisetal2012} {Bountis, T. \& Skokos, H.} [2012] {\it Complex Hamiltonian Dynamics} (Springer-Verlag Berlin Heidelberg).

\bibitem[{{Cannon}(1932)}] {Cannon1932} {Cannon, W. B.} [1932] {\it The Wisdom of the Body} (W. W. Norton \& Co, New York).

\bibitem[{{Capuzzo-Dolcetta et al}(2007)}] {CapuzzoDolcettaetal2007} {Capuzzo-Dolcetta, R., Leccese, L., Merritt, D. \& Vicari, A.} [2007] ``Self-consistent models of cuspy triaxial galaxies with dark matter halos,'' {\it The Astrophysical Journal}, {\bf 666}, 1, 165.

\bibitem[{{Cinquin \& Demongeot}(2002)}] {Cinquinetal2002} {Cinquin, O. \& Demongeot, J.} [2002] ``Positive and negative feedback: striking a balance between necessary antagonists,'' {\it J. Theoret. Biol.}, {\bf 216}, 229--241. 

\bibitem[{{Demongeot et al}(2012)}] {Demongeotetal2012A} {Demongeot, J., Noual, M. \& Sen\'{e}, S.} [2012A] ``Combinatorics of Boolean automata circuits dynamics,'' {\it Discrete Applied Mathematics}, {\bf 160}, 398--415.

\bibitem[{{Demongeot \& Waku}(2012)}] {Demongeotetal2012B} {Demongeot, J. \& Waku, J.} [2012B] ``Robustness in biological regulatory networks: I \& II,'' {\it Comptes Rendus Math\'{e}matique}, {\bf 350}, 221--228 \& 289--298.

\bibitem[{{Demongeot et al}(2011)}] {Demongeotetal2011} {Demongeot, J., Elena, A., Noual, M. \& Sen\'{e}, S.} [2011] ``Random Boolean networks and attractors of their intersecting circuits,'' {\it Proc. IEEE AINA'} {\bf 11} {\it \& BLSMC} {\bf 11} {\it and Proc. IEEE Proceedings} (Piscataway), pp. 483--487.

\bibitem[{{Demongeot et al}(2007)}] {Demongeotetal2007} {Demongeot, J., Glade, N. \& Forest, L.} [2007] ``Li\'{e}nard systems and potential-Hamiltonian decomposition: I \& II,'' {\it Comptes Rendus Math\'{e}matique}, {\bf 344}, 121--126 \& 191--194.

\bibitem[{{Della Dora et al}(2003)}] {Dellaetal2003} {Della Dora, J., Mirica-Ruse, M. \& Tournier, E.} [2003] ``Hybrid systems and hybrid computation,'' {\it Numerical Algorithms}, {\bf 33}, 203--213.

\bibitem[{{Eisenfeld \& de Lisi}(1985)}] {EisenfeldanddeLisi} {Eisenfeld, J. \& de Lisi, C.} [1985] ``On conditions for qualitative instability of regulatory circuits with application to immunological control loops,'' {\it Mathematics and Computers in Biomedical Applications, edited by Eisenfeld, J. and de Lisi, C. eds.}, Elsevier, Amsterdam, 39--53.

\bibitem[{{Goles}(1985)}] {Goles1985} {Goles, E.} [1985] ``Dynamics of positive automata networks,'' {\it Theor. Comput. Sci.}, {\bf 41}, 19--32.

\bibitem[{{Lichtenberg \& Lieberman}(1983)}] {Lichtenbergetal1983} {Lichtenberg, A. J. \& Lieberman, M. A.} [1983] {\it Regular and stochastic motion} (Springer-Verlag, New York Heidelberg Berlin, Applied Mathematical Sciences).

\bibitem[{{Manos \& Athanassoula}(2011)}] {Manosetal2011} {Manos, T. \& Athanassoula, E.} [2011] ``Regular and chaotic orbits in barred galaxies - I. Applying the {SALI/GALI} method to explore their distribution in several models,'' {\it Mon. Not. R. Astron. Soc.}, {\bf 415}, 629--642.

\bibitem[{{Manos et al}(2008)}] {Manosetal2008} {Manos, T., Skokos, Ch., Athanassoula, E. \& Bountis, T.} [2008] ``Studying the global dynamics of conservative dynamical systems using the SALI chaos detection method,'' {\it Nonlinear Phenomena in Complex Systems}, {\bf 11}, 2, 171--176.

\bibitem[{{Oseledec}(1968)}] {Oseledec1968} {Oseledec, V. I.} [1968] ``A multiplicative ergodic theorem and Lyapunov characteristic numbers for dynamical systems,'' {\it Trans. Moscow Math. Soc.}, {\bf 19}, 197--231.

\bibitem[{{Pesin}(1977)}] {Pesin1977} {Pesin, Y. B.} [1977] ``Characteristic Lyapunov exponents and smooth ergodic theory,'' {\it Russian Math. Surveys}, {\bf 32}, 55--114.

\bibitem[{{Plahte et al}(1995)}] {Plahte1995} {Plahte, E., Mestl, T. \& Omholt, S. W.} [1995] ``Feedback loops, stability and multistationarity in dynamical systems,'' {\it J. Biol. Syst.}, {\bf 3}, 409--413.

\bibitem[{{Richard \& Comet}(2011)}] {Richardetal2011} {Richard, A. \& Comet, J.-P.} [2011] ``Stable periodicities and negative circuits in differential systems,'' {\it J. Math Biol.}, {\bf 63}, 593--600.

\bibitem[{{Richard}(2011)}] {Richard2011} {Richard, A.} [2011] ``Local negative circuits and fixed points in non-expansive Boolean networks,'' {\it Discrete Applied Mathematics}, {\bf 159}, 1085--1093.

\bibitem[{{Sarraille \& Myers}(1994)}] {Sarrailleetal1994} {Sarraille, J. J. \& Myers, L. S.} [1994] ``FD3: A program for measuring fractal dimension,'' {\it Educational and Psychological Measurement}, {\bf 54}, 1, 94--97.

\bibitem[{{Skokos et al}(2004)}] {Skokosetal2004} {Skokos, Ch., Antonopoulos, Ch., Bountis, T. \& Vrahatis, M. N.} [2004] ``Detecting order and chaos in Hamiltonian systems by the SALI method,'' {\it J. Phys. A: Math. \& Gen.}, {\bf 37}, 6269--6284.

\bibitem[{{Skokos}(2001)}] {Skokos2001} {Skokos, Ch.} [2001] ``Alignment Indices: A new, simple method for determining the ordered or chaotic nature of orbits,'' {\it J. Phys. A: Math. \& Gen.}, {\bf 34}, 10029--10043.

\bibitem[{{Skokos et al}(2002)}] {Skokosetal2002} {Skokos, Ch., Antonopoulos, Ch., Bountis, T. \& Vrahatis, M. N.} [2002] ``Smaller Alignment Index (SALI): Detecting order and chaotic motion in conservative dynamical systems,'' {\it Proc. $4^{\mbox{th}}$ GRACM Congress on Computational Mechanics (GRACM 2002)}, (Patras, Greece), pp. 27--29.

\bibitem[{{Skokos et al}(2003)}] {Skokosetal2003} {Skokos, Ch., Antonopoulos, Ch., Bountis, T. \& Vrahatis, M. N.} [2003] ``How does the Smaller Alignment Index (SALI) distinguish order from chaos?,'' {\it Prog. Theor. Phys. Supp.}, {\bf 150}, 439--443.

\bibitem[{{Skokos et al}(2007)}] {Skokosetal2007} {Skokos, Ch., Bountis, T. C. \& Antonopoulos, Ch.} [2007] ``Geometrical properties of local dynamics in Hamiltonian systems: The Generalized Alignment Index (GALI) method,'' {\it Physica D}, {\bf 231}, 30--54.

\bibitem[{{Skokos et al}(2008)}] {Skokosetal2008} {Skokos, Ch., Bountis, T. C. \& Antonopoulos, Ch.} [2008] ``Detecting chaos, determining the dimensions of tori and predicting slow diffusion in Fermi­-Pasta-Ulam lattices by the Generalized Alignment Index method,'' {\it Eur. Phys. J. Special Topics}, {\bf 165}, 5--14.

\bibitem[{{Skokos}(2010)}] {Skokos2010} {Skokos, Ch.} [2010] ``The Lyapunov Characteristic Exponents and their computation,'' {\it Lect. Notes Phys.}, {\bf 790}, 63--135.

\bibitem[{{Snoussi}(1998)}] {Snoussi1998} {Snoussi, E. H.} [1998] ``Necessary conditions for multistationarity and stable periodicity,'' {\it J. Biol. Syst.}, {\bf 6}, 3--9.
 
\bibitem[{{Soule}(2003)}] {Soule2003} {Soule, C.} [2003] ``Graphic requirements for multistationarity,'' {\it ComPlexUs}, {\bf 1}, 123--122.

\bibitem[{{Sparrow}(1982)}] {Sparrow1982} {Sparrow, C.} [1982] {\it The Lorenz Equations: Bifurcations, Chaos, and Strange Attractors} (Springer-Verlag, New York Heidelberg Berlin, Applied Mathematical Sciences).

\bibitem[{{Thomas \& Nardone}(2009)}] {ThomasandNardone2009} {Thomas, R. \& Nardone, P.} [2009] ``A further understanding of phase space partition diagrams,'' {\it Int. J. Bifurcat. Chaos Appl. Sci. Eng.}, {\bf 19}, 3, 785--804.

\bibitem[{{Thomas \& Kaufman}(2001)}] {ThomasandKaufman2001} {Thomas, R. \& Kaufman, M.} [2001] ``Multistationarity, the basis of cell differentiation and memory: I. Structural conditions of multistationarity and other nontrivial behavior,'' {\it Chaos}, {\bf 11}, 1, 170--179.

\bibitem[{{Thomas}(1999)}] {Thomas1999} {Thomas, R.}.[1999] ``Deterministic chaos seen in terms of feedback circuits: analysis, synthesis, ``labyrinth chaos'','' {\it Int. J. Bifurcation Chaos Appl. Sci. Eng.}, {\bf 9}, 1889--1905.

\bibitem[{{Thomas}(1981)}] {Thomas1981} {Thomas, R.} [1981] ``On the relation between the logical structure of systems and their ability to generate multiple steady states or sustained oscillations,'' {\it Springer series in Synergetics}, {\bf 9}, 180--193.

\bibitem[{{Thomas \& R\"{o}ssler}(2006)}] {Thomasetal2006} {Thomas, R. \& R\"{o}ssler, O. E.} [2006] ``Gen\`{e}se de formes,'' {\it Revue des Questions Scientifiques}, {\bf 177}, 271--287.

\bibitem[{{Thomas \& Kaufman}(2005)}] {Thomasetal2005} {Thomas, R. \& Kaufman, M.} [2005] ``Frontier diagrams: Partition of phase space according to the signs of the eigenvalues or to the sign patterns of the circuits,'' {\it Int. J. Bifurcat. Chaos}, {\bf 15}, 3051--3074.

\bibitem[{{Tsypkin}(1958)}] {Tsypkin1958} {Tsypkin, Ya. Z.} [1958] {\it Theorie der Relaissysteme der Automatische Regelung} (R. Oldenbourg Verlag Munchen, Verlag Technik, Berlin, Germany).

\bibitem[{{Voglis et al}(1999)}] {Voglisetal1999} {Voglis, N., Contopoulos, G. \& Efthymiopoulos, C.} [1999] ``Detection of ordered and chaotic motion using the dynamical spectra,'' {\it Celestial Mechanics and Dynamical Astronomy}, {\bf 73}, 211--220.

\end{thebibliography}

\end{document}